\documentclass[%
  twoside,
  floatfix,
  reprint,
  amsmath,amssymb,
  aps,
  pra,
  nofootinbib,
  showpacs,
  superscriptaddress,
  a4paper
]{revtex4-1}

\usepackage{ORCIDinREVTeX}
\usepackage{graphicx}% 
\usepackage{subfigure}
\usepackage{enumitem}
\usepackage{subfigure}

\usepackage{tabularx}
\usepackage{booktabs}
\usepackage{multirow}
\usepackage{color}
\usepackage{titlesec}

\usepackage{array}

\usepackage[utf8]{inputenc}
\usepackage[T1]{fontenc}

\usepackage{lipsum}

\graphicspath{{Figures/}{}}

\usepackage[
centering, includefoot,
text={7.1in,10.2in},
total={6.3in,8.75in},
top=0.8in, left=0.62in,
]{geometry}

\usepackage[
  bookmarks=false,
  colorlinks,
  linkcolor=blue,
  urlcolor=blue,
  citecolor=blue,
  plainpages=false,
  pdfpagelabels,
  final,
  breaklinks=true
]{hyperref}
\hypersetup{
pdftitle={Principal frequency of an ultrashort laser pulse}, 
pdfauthor={Authors}
}
\usepackage{natbib}
\makeatletter \def\NAT@def@citea{\def\@citea{\NAT@separator\,}} \makeatother

\newcommand{\reffig}[1]{Fig.~\ref{#1}}

\usepackage{physics}

  \newcommand{\sinc}{\mathrm{sinc}}
 
\newcommand{\rmd}{\mathrm{d}}

\newcommand{\eref}[1]{Eq.~(\ref{#1})}

\newcommand{\nhphantom}[1]{\sbox0{#1}\hspace{-\the\wd0}}
\begin{document}

%\preprint{APS/123-QED}

\title{Principal frequency of an ultrashort laser pulse}% Force line breaks with \\

\author{Enrique G. Neyra}
\affiliation{CIOp: Centro de Investigaciones \'Opticas, CONICET-CICBA-UNLP, Camino Centenario y 506, M.B. Gonnet (1897), Pcia. Bs. As., Argentina.}

\author{Pablo Vaveliuk}
\affiliation{CIOp: Centro de Investigaciones \'Opticas, CONICET-CICBA-UNLP, Camino Centenario y 506, M.B. Gonnet (1897), Pcia. Bs. As., Argentina.}

\author{Emilio Pisanty}
\orcid{0000-0003-0598-8524}
 \affiliation{ICFO -- Institut de Ciencies Fotoniques, The Barcelona Institute of Science and Technology, 08860 Castelldefels (Barcelona)}
 \affiliation{Max Born Institute for Nonlinear Optics and Short Pulse Spectroscopy, Max-Born-Stra{\ss}e 2A, Berlin 12489, Germany}

\author{Andrew S. Maxwell}
\orcid{0000-0002-6503-4661}
\affiliation{ICFO -- Institut de Ciencies Fotoniques, The Barcelona Institute of Science and Technology, 08860 Castelldefels (Barcelona)}
\affiliation{Department of Physics \& Astronomy, University College London, Gower Street, London WC1E 6BT, United Kingdom}

\author{Maciej Lewenstein}
\orcid{0000-0002-0210-7800}
 \affiliation{ICFO -- Institut de Ciencies Fotoniques, The Barcelona Institute of Science and Technology, 08860 Castelldefels (Barcelona)}
 \affiliation{ICREA, Passeig de Llu\'is Companys, 23, 08010 Barcelona, Spain}

\author{Marcelo F. Ciappina}
\orcid{0000-0002-1123-6460}
 \email{marcelo.ciappina@gtiit.edu.cn}
 \affiliation{ICFO -- Institut de Ciencies Fotoniques, The Barcelona Institute of Science and Technology, 08860 Castelldefels (Barcelona)}
\affiliation{Physics Program, Guangdong Technion -- Israel Institute of Technology, Shantou, Guangdong 515063, China}
\affiliation{Technion -- Israel Institute of Technology, Haifa, 32000, Israel}

%\affiliation{%


%\affiliation{
%\affiliation{
%\affiliation{%



\begin{abstract}
We introduce an alternative definition of the main frequency of an ultrashort laser pulse, the principal frequency $\omega_P$. This parameter is complementary to the most accepted and widely used carrier frequency $\omega_0$. Given the fact that these ultrashort pulses, also known as transients, have a temporal width comprising only few cycles of the carrier wave, corresponding to a spectral bandwidth $\Delta\omega$ covering several octaves, $\omega_P$ describes, in a more precise way, the dynamics driven by these sources. We present examples where, for instance, $\omega_P$ is able to correctly predict the high-order harmonic cutoff independently of the carrier envelope phase. This is confirmed by solving the time-dependent Schr\"odinger equation in reduced dimensions, supplemented with the time-analysis of the quantum spectra, where it is possible to observe how the sub-cycle electron dynamics is better described using $\omega_P$. The concept of $\omega_P$, however, can be applied to a large variety of scenarios, not only within the strong field physics domain.

\end{abstract}

\maketitle

%\tableofcontents

\section{Introduction}

During the past two decades we have been witness to a constant development of a varied set of ultrashort laser pulses, with temporal widths well below two optical cycles. In general, the main objective of these sources is the temporal study of diverse physical phenomena on their natural scale. The techniques that have been developed to scrutinize dynamics on this territory are based on delicate control of strong-field laser-atom interactions and configure the core of what is known as attosecond spectroscopy. The spectral range of those pulses is very broad, covering both the THz~\cite{hwang2015review,nanni2015terahertz,reimann2007table}, and infrared and visible,~\cite{brabec2000intense,krogen2017generation,hwang2019generation,lu2018sub} as well as the XUV~\cite{schultze2014attosecond,goulielmakis2008single,sansone2006isolated,Krausz2009,hassan2016} regions of the electromagnetic spectrum.

The driving sources described above also allow the coherent control of various quantum systems, particularly the standard two-level system, widely used as a toy model for different physical processes. The precise and high-speed manipulation over the population of quantum states  has important applications in, e.g., quantum information and spintronics, among others~\cite{Kampfrath2011}. In addition, the generation of intense ultrashort laser pulses in the few-cycle, single-cycle and sub-cycle domains has enabled the study of strongly non-linear light-matter interactions and given rise to novel and fascinating phenomena~\cite{Popmintchev2012}. Maybe the most important example are the so-called isolated attosecond pulses (IAPs). These sources are the workhorse to tackle the dynamics of electrons under strong fields in its natural temporal, attosecond, scale~\cite{Krausz2009, Corkum2007}. IAPs are obtained from high-harmonic generation (HHG) using a variety of spectral postprocessing approaches~\cite{Ivanov2014,Lewenstein1994,Amini2019,LHuillier1993,Corkum1993}. 

HHG is an extremely non-linear optical process in which a strong laser field interacts with atoms, molecules and, recently, bulk materials, and drives
the production of high-frequency ultrashort bursts of coherent electromagnetic radiation~\cite{Krausz2009, Corkum2007}. This emission possesses a set of distinct features, namely, (i) a steadying decrease of the first harmonics of the driving field yield, (ii) a broad plateau, that can cover up to thousands of harmonic orders of the original driving field, and (iii) a cutoff, where the spectrum suddenly terminates. The underlying physics of the HHG in atoms and molecules can be traced out from a sequence involving three steps, which can be summarized as follows: (i) the laser ionizes the target via tunnel ionization, (ii) the released electron travels in the laser continuum gaining kinetic energy and, when the laser electric field reverses its direction, (iii) the electron returns back to the parent ion, where it recombines, releasing its kinetic energy as a high energy photon~\cite{Corkum1993}.

The HHG phenomenon can be modelled using a wide range of approaches, from classical-based schemes~\cite{Corkum1993, Lewenstein1994} to intensive numerical computations involving the numerical solution of the time-dependent Schr\"odinger equation (TDSE), both in one or several spatial dimensions~\cite{Bauer2006}. Yet, the quantitative schemes that most closely follow the overall intuition are the so-called quasi-classical methods, with the Strong-Field Approximation (SFA) being the most prominent exponent~\cite{Lewenstein1994,Amini2019}. Here the emission amplitude key ingredient is a path-integral sum over discrete emission events. Invoking the SFA, the well known 3.17-law, which correctly predicts the HHG cutoff law, can be easily obtained~\cite{Emilio2020}. The SFA has been applied to a large variety of strong field processes with undeniable success (for a recent historical review see~\cite{Amini2019}).

The synthesis of ultrashort pulses has advanced quickly in recent years. Particularly, the generation of high-energy single- and sub-cycle IR laser pulses has been experimentally verified, through the combination and manipulation of the spectral content of laser sources of different wavelengths. The current generation of these pulses possesses tremendous technological challenges, due to, amongst other difficulties, the synchronization of their different sources with sub-fs temporal resolution~\cite{poppe2001few,cox2012pulse,krauss2010synthesis,manzoni2015coherent,huang2011high,hassan2012invited,wirth2011synthesized,Fattahi2014}. Recently, the generation of a 53-attosecond X-ray pulse was demonstrated using HHG in noble gases driven by a mid-infrared few-cycle laser source~\cite{Li2017}. 

When working with ultrashort laser pulses in the few-cycle regime, it is well known that the so-called carrier-envelope phase (CEP), $\phi$, plays an instrumental role in the resulting laser-matter interaction processes driven by those sources. This is because the pulse envelope experiences appreciable changes within an optical cycle of the carrier wave. In this way, for instance, the maximum field amplitude of a sine-like pulse, typically characterized by $\phi=0$, is largely different than that of a cosine-like pulse, where $\phi=\pi/2$. Furthermore, it has been demonstrated 
that not only the CEP, but also the pulse width is relevant in certain strong field processes~\cite{de1998phase,brabec1997nonlinear,li2010carrier,venzke2018}.

In this work we introduce an alternative definition of the main frequency of an ultrashort laser pulse. This new parameter, which we name principal frequency $\omega_P$, appears to be much more appropriate than the standard definition, the carrier frequency $\omega_0$, to correctly characterize the interaction of these pulses with matter. Using $\omega_P$ as the frequency that drives the dynamics, we are able to give a better interpretation of previously published results, as well as provide more reliable predictions of strong field processes outcomes. Its definition is based on a particular way to \textit{weight} the spectral content of the laser pulse electric field and it is adequately justified if we resort to the particle nature of light, i.e., if we consider that light is composed of light-quanta (photons).

This article is organized as follows. In Section~\ref{wp}, we present the mathematical foundations of the principal frequency $\omega_P$. We present a set of examples based on different definitions of the laser electric field. We show how $\omega_P$ varies as a function of the bandwidth of the pulses for three archetypal cases. Furthermore we show that there exists a correlation between the positions of the maxima and minima of the laser pulse's electric field with the principal period, defined as $T_P=2\pi/\omega_P$. In Section~\ref{interaction} we use the definition of $\omega_P$ to characterize the HHG spectra of an atom driven by a series of few-cycle laser pulses. For the computation of the HHG spectra, we use both quantum mechanical and classical approaches. These two complementary schemes allow us to disentangle the underlying physics of the HHG process. We end our contribution in Section~\ref{conclusions} presenting our conclusions together with a brief outlook. Atomic units are used throughout the article unless otherwise stated.

\section{Principal frequency}
\label{wp}
\subsection{Definition} 
 
The most accepted definition of carrier frequency, $\omega_0$, of an ultrashort laser electric field pulse $E(t)$ is the central frequency of the modulus of its Fourier transform $|E(\omega)|$, if the spectrum is symmetric. However, if the spectral content is more complex, $\omega_0$ results from an integral over the density distribution $\rho(\omega)= S(\omega)$, where  $S(\omega)=|E(\omega)|^2$ is the spectral power, i.e.~

\begin{equation}\label{e1}
\omega_0=\frac{\int_{-\infty}^{\infty}\omega S(\omega) \rmd\omega}{\int_{-\infty}^{\infty}S(\omega)\rmd\omega}.
\end{equation}

If we think of laser pulses born in an optical cavity as a frequency comb with $n$ longitudinal modes, each with individual frequencies $\omega_i$, considering that the integral in the denominator $\int_{-\infty}^{\infty}S(\omega) \rmd\omega$ defines the total energy of the pulse $E_p$, we can write this term as a sum over $n$ modes as $\sum_{i}S_i(\omega_i)=E_p$. Here, $S_i(\omega_i)$ is the energy of the $i-$mode, which can be written in terms of the number of photons as $S_i(\omega_i)=\hbar\omega_in_i$. Therefore, we can interpret the above definition as an average over the energy $S_i(\omega_i)$ of each mode:

\begin{equation}
\label{e2}
\omega_0=\frac{\sum_{i}\omega_i S_i(\omega_i)}{\sum_{i}S_i(\omega_i)}.
\end{equation}

Let us suppose, for example, that we have two modes with frequencies $\omega_1$ and $\omega_2$ in the frequency comb, with energies $S_1(\omega_1)=S_2(\omega_2)$. Independently of the values of the frequencies $\omega_1$ and $\omega_2$, the carrier frequency $\omega_0$ can be easily calculated as $\omega_0=(\omega_1+\omega_2)/2$, even if $\omega_1<<\omega_2$. In other words, if the spectrum of the pulse $S(\omega)$ is symmetrical, the carrier frequency is independent of the bandwidth of the pulse.  

Nothing prevent us from given different \textit{weights} to  $\omega_1$ and $\omega_2$ in the average. For instance, one reasonable choice would be to 
suppose that more energetic photons with higher frequencies have more \textit{weight} in the average. In this way, the new density function takes the following form: $\rho(\omega)=\omega S(\omega)$.

%Considering the integral in the denominator $\int_{-\infty}^{\infty}S(\omega) \rmd\omega$ defines the total number of photons $\sum_{i}n_i=N$, we can interpret the above %definition as an average over the photon energies:
%\begin{equation}
%\label{e2}
%\hbar\omega_0=\frac{\sum_{i}\hbar\omega_i n_i}{\sum_{i}n_i}.
%\end{equation}

%Let us construct a different explanation. The definition above establishes that every photon has the same \textit{weight} in the average of \eref{e2}, i.e., there is no difference between $n_0$ photons with energy $\omega_1$ and $n_0$ photons with energy $\omega_2$, where, e.g., $\omega_1<\omega_2$. But, on the contrary, it could be natural to assume that a more energetic photon has a greater weight in the integral. Therefore, we can change the density function $\rho(\omega)$ by an \textit{energy} distribution, i.e.~to use $\hbar\omega_i n_i$ instead of the number of photons $n_i$ in \eref{e2}. In this way, the new density function takes the following form: $\rho(\omega)=\omega S(\omega)$.  

In this way, our principal frequency $\omega_P$ results:
\begin{eqnarray}
\label{e3}
\omega_P&=&\frac{\int_{-\infty}^{\infty}\omega^2 S(\omega) \rmd\omega}{\int_{-\infty}^{\infty}\omega S(\omega)  \rmd\omega}.
\end{eqnarray}
From the above definition we can observe that $\omega_P$ gives more weight to photons with greater frequencies (higher energies). This is so because the actual definition of the new density function $\rho(\omega)$. Let us now see what this means in the temporal domain.

The electric field of an ultrashort laser pulse can be written as $E(t)=f(t)e^{i\omega_0 t}e^{i\phi}$, where $f(t)$, $\omega_0$ and $\phi$ are the pulse envelope, the carrier frequency and the so-called carrier-enveloped phase (CEP), respectively. If we take $\mathrm{Re}[E(t)]=0$, we can find the \textit{zeros} of $E(t)$, i.e.~the times where $E(t)=0$. As is well known, these zeros are located at $n\pi/\omega_0$ and $(2n+1)\pi/2\omega_0$ for sine-like and cosine-like pulses, respectively, and are spaced by $\pi/\omega_0$. Here, $n=0,1,\ldots$. But what happens with the position of the maxima and minima of the field $E(t)$? 
For long pulses, i.e.~when the temporal width $\tau$ is $\tau \gg T_0$, where $T_0=2\pi/\omega_0$ is the carrier period, the envelope in the central region varies slowly and the maxima and minima are spaced by $T_0/2$. However, if $\tau$ is of the order of $T_0$, i.e.~$\tau \sim T_0$, the situation changes considerably. For this case, it is easy to see that the maxima and minima of the field $E(t)$ depends now both on the envelope $f(t)$ and the argument of the phase $e^{i\omega_0 t}e^{i\phi}$. This simple conclusion is important, if we consider that the interaction of ultrashort pulses with matter is dominated by these maxima and minima, and not by the \textit{zeros} of $E(t)$. 

In the next section we show how $\omega_P$ is correlated with the positions of the maxima and minima of the electric field $E(t)$, for different cases, each of them with different spectral content.

\begin{figure*}
{\setlength{\tabcolsep}{-0.5mm}
\newlength{\spacing}
\setlength{\spacing}{-6mm}
\begin{tabular}{rl}
\subfigure{\label{thpulses-a} %
\includegraphics[width=0.5\textwidth]{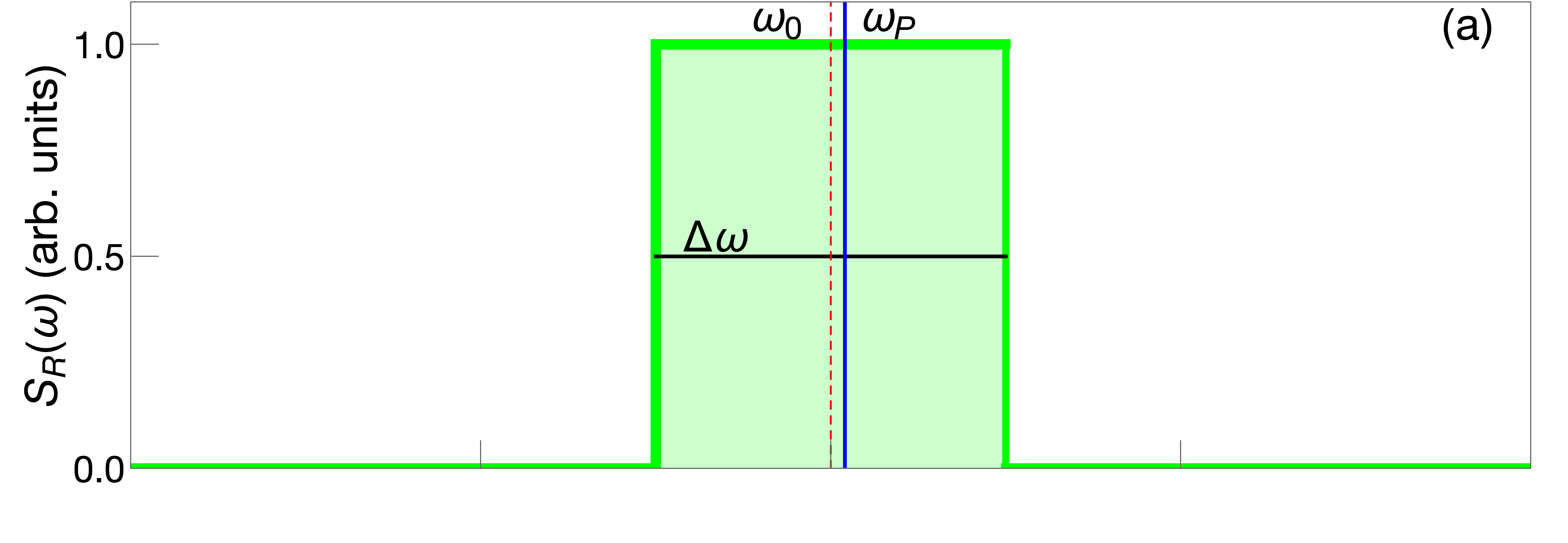} } & 
\subfigure{\label{thpulses-d} %
\includegraphics[width=0.5\textwidth]{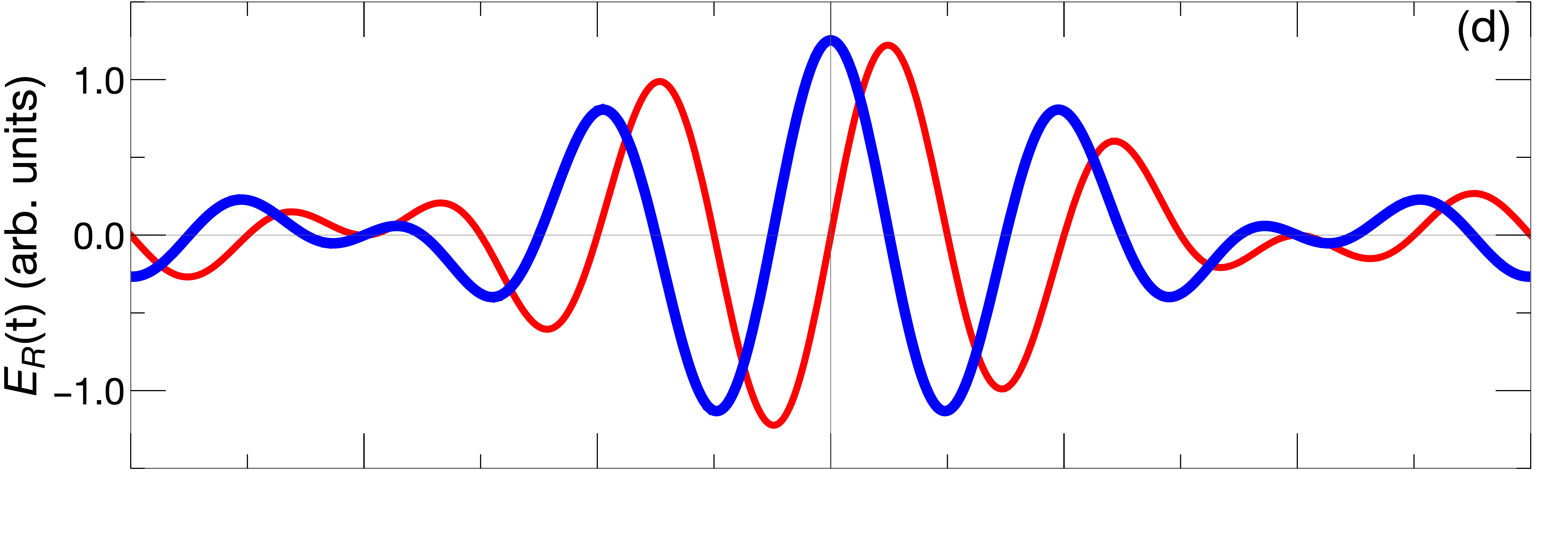} }\\[\spacing]
\subfigure{\label{thpulses-b} %
\includegraphics[width=0.5\textwidth]{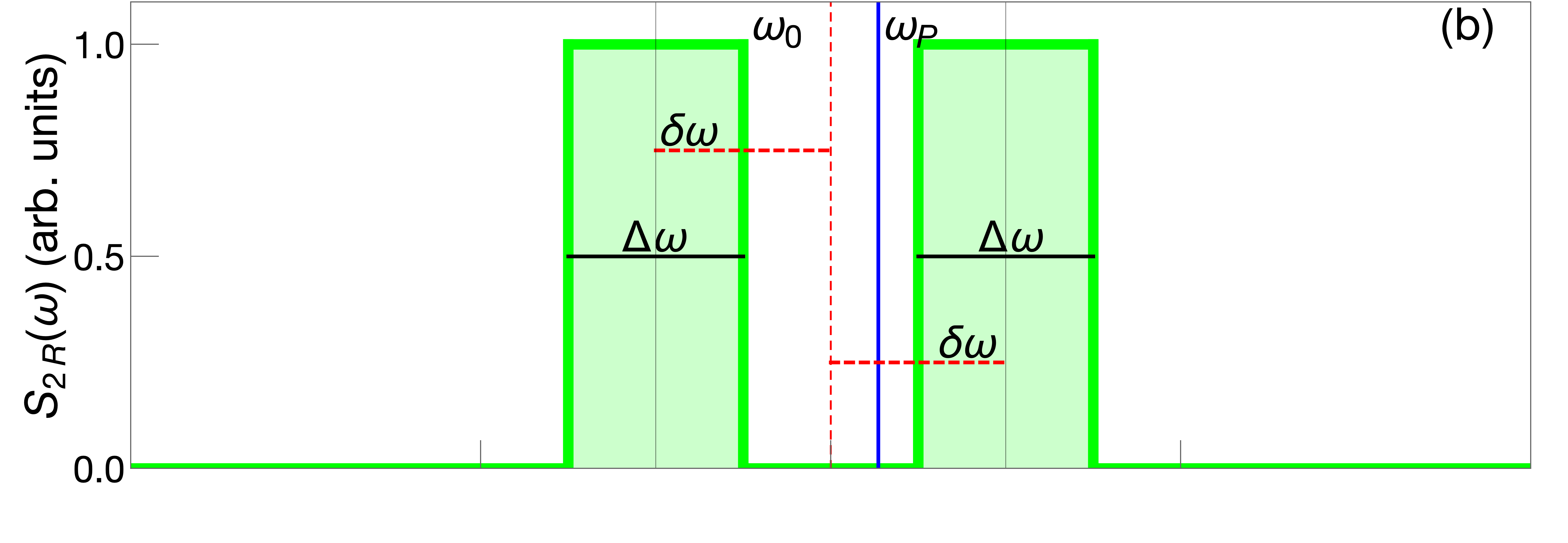} } & 
\subfigure{\label{thpulses-e} %
\includegraphics[width=0.5\textwidth]{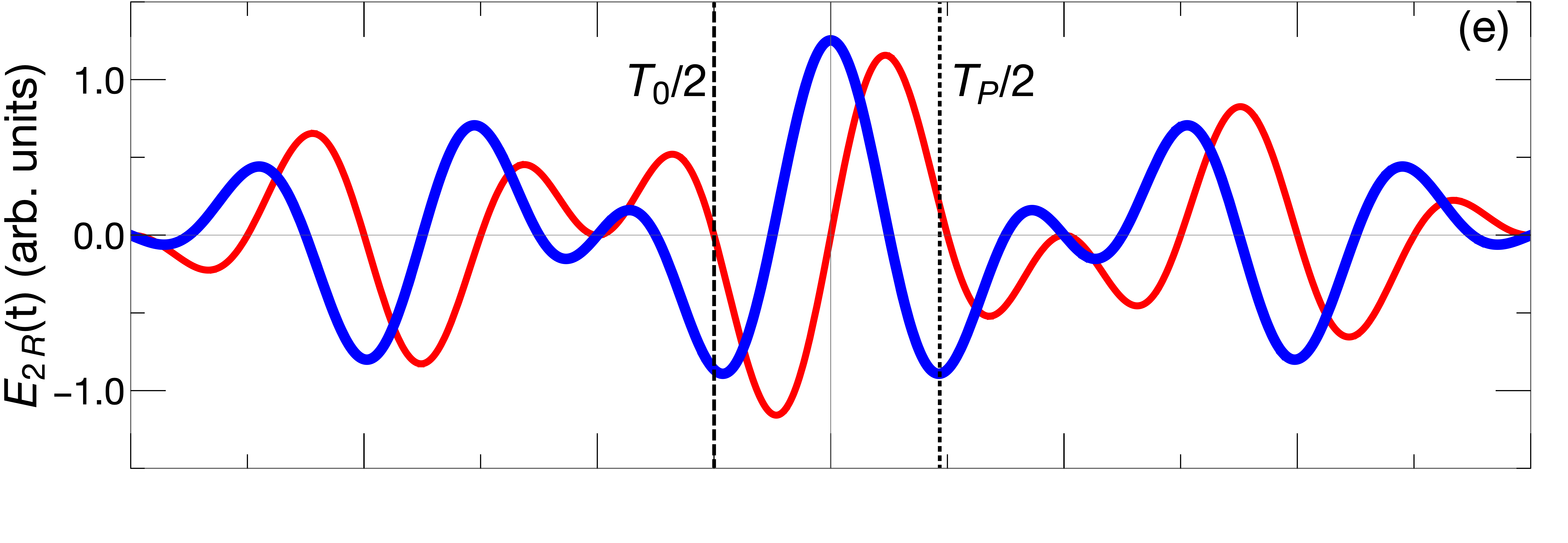} }\\[\spacing]
\subfigure{\label{thpulses-c} %
\includegraphics[width=0.5\textwidth]{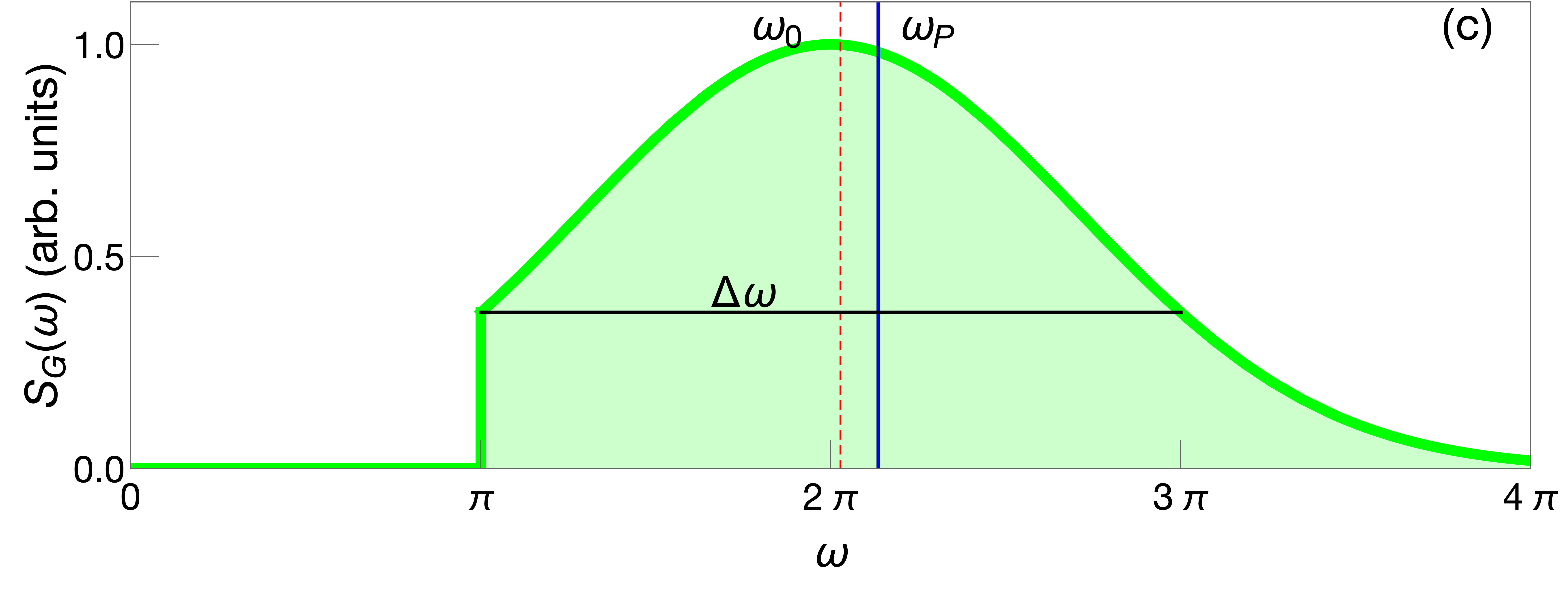} } & 
\subfigure{\label{thpulses-f} %
\includegraphics[width=0.5\textwidth]{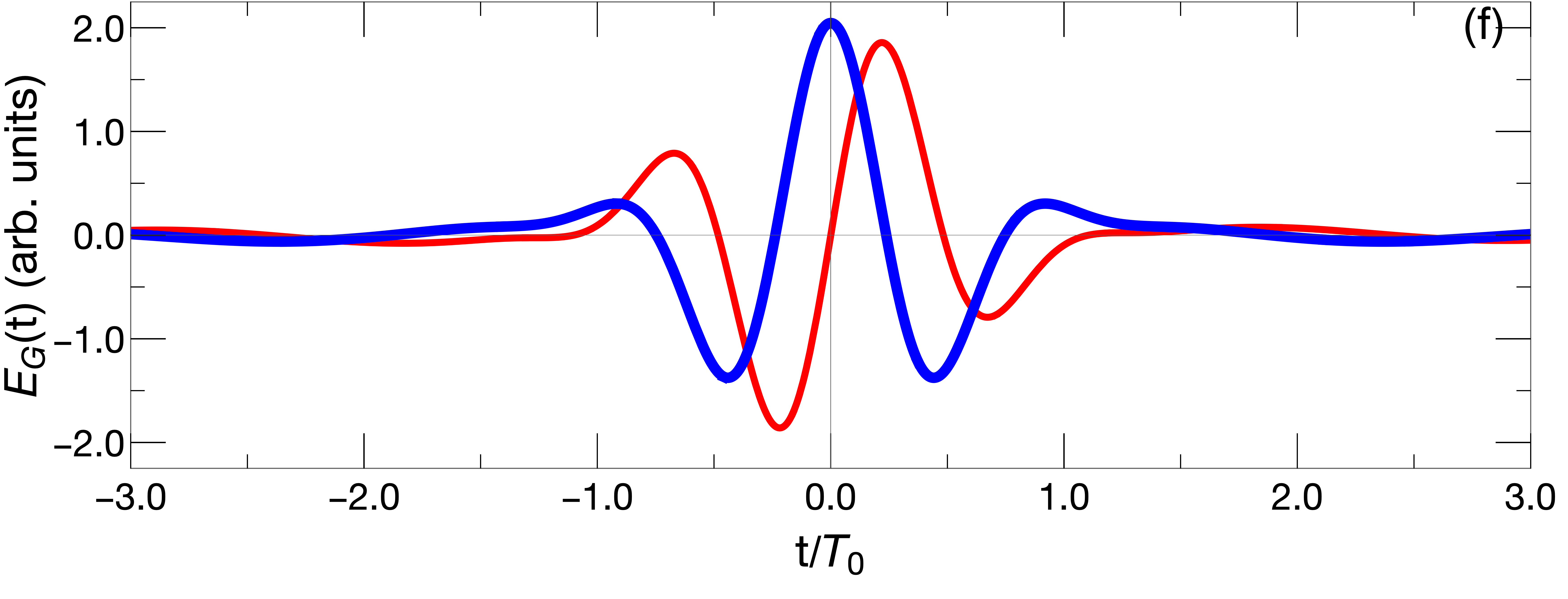} }
\end{tabular}
}
\caption{(a)-(c) Spectral power $S(\omega)$ of the fields $E_R(\omega)$, $E_{2R}(\omega)$ and $E_G(\omega)$ for $E_0=1$. All the cases are centered at the same frequency, $\omega_1=2\pi$. The spectral bandwidth of both $E_R(\omega)$ and $E_G(\omega)$ is $\Delta\omega=\pi$, meanwhile for $E_{2R}(\omega)$ each `sub-spectrum' has a bandwidth $\Delta\omega=\pi/2$. In (b) $\delta\omega=\pi/2$. The blue solid (red dashed) represents the $\omega_P$ ($\omega_0$) frequency (see the text for details). (d)-(f) normalized time-dependent fields $E_R(t)$, $E_{2R}(t)$ and $E_G(t)$. Thick blue (thin red) line defines a sine(cosine)-like pulse. In (e) the dashed (dotted) line corresponds to the temporal distance between two consecutive zeros (maximum and minimum) of the field for the $\sin$($\cos$)-like pulses, i.e.~$T_0/2$ and $T_P/2$, respectively (see the text for details).
}
\label{thpulses}
\end{figure*}

\subsection{Analysis} 

We start our analysis by considering the following fields in the spectral domain:
\begin{subequations}
\begin{eqnarray}
E_R(\omega) & = & E_0\,\mathrm{rect}\left(\frac{\omega-\omega_1}{\Delta\omega}\right) \label{erw}\\  
E_{2R}(\omega) & = & E_0\,\mathrm{rect}\left(\frac{\omega-\omega_1+\delta\omega}{\Delta\omega}\right)\nonumber\\
&&+E_0\,\mathrm{rect}\left(\frac{\omega-\omega_1-\delta\omega}{\Delta\omega}\right) \label{e2rw}\\
E_G(\omega) & = &E_0\,e^{-\left(\frac{\omega-\omega_1}{\Delta\omega}\right)^2}\mathrm{rect}\left(\frac{\omega-5\omega_1}{9\omega_1}\right)
\label{egw},
\end{eqnarray}
\end{subequations}
where $E_0$ is the peak field strength and the function $\mathrm{rect}(x/x_0)$ is the so-called rectangle function defined as:
\begin{equation}
 \mathrm{rect}(x/x_0)= 
     \begin{cases}
       \text{1} &\quad\text{if}\; |x|\leq x_0/2\\
       \text{0} &\quad\text{if}\; |x| > x_0/2. \\
     \end{cases}
\end{equation}

The spectra of Eqs.~(\ref{erw}) and (\ref{egw}) are centered at the frequency $\omega_1$, meanwhile \eref{e2rw} is the sum of two carrier waves with different frequencies, $\omega_1 +\delta\omega$ and $\omega_1 -\delta\omega$. In all the cases $\Delta\omega$ characterizes their respective spectral bandwidths. 

By taking the Fourier transform we find the electric fields in the temporal domain,i.e.~
\begin{subequations}
\begin{eqnarray}
E_R(t)&=&E_0\,\frac{\Delta\omega}{\sqrt{2\pi}}e^{i\omega_1 t} \sinc\left(\frac{\Delta\omega\,t}{2}\right) \label{ert}\\  
E_{2R}(t)&=&E_0\,\frac{\Delta\omega}{\sqrt{2\pi}}e^{i t(\omega_1-\delta\omega)}\sinc\left(\frac{\Delta\omega\, t}{2}\right)\nonumber\\ 
&\times& (1+e^{2i\delta\omega t} ) \label{e2rt} \\
E_G(t)&=&E_0\,\frac{i\Delta\omega\, e^{-(\frac{\Delta\omega t}{2})^2}e^{i\omega_1 t}U(t)}{2\sqrt{2}} \label{egt},
%\\
%U(t)=(erfi(\frac{\Delta\omega t}{2}-i\frac{\omega_0}{2\Delta\omega})-erfi(\frac{\Delta\omega t}{2}+i\frac{17\omega_0}{2\Delta\omega}))
\end{eqnarray}
\end{subequations}
where $U(t)$ is given by
\begin{eqnarray}
U(t)&=&\mathrm{erfi}\mathopen{}\left(\frac{\Delta\omega t}{2}-i\frac{\omega_1}{2\Delta\omega}\right)\mathclose{}\nonumber\\
&&-\mathrm{erfi}\mathopen{}\left(\frac{\Delta\omega t}{2}+i\frac{17\omega_1}{2\Delta\omega}\right)\mathclose{},
\end{eqnarray}
and $\sinc$ and $\mathrm{erfi}$ are the $\sinc$ function $\sinc(x)=\sin(x)/x$ and the imaginary error function, respectively.

%\begin{figure*}[ht!]
%\includegraphics[width=0.75\textwidth]{Fig1.0.png}% Here is how to import EPS art
%\caption{\label{fig2} A figure caption. The figure captions are
%automatically numbered.}
%\end{figure*}

Note that the spectra of the first two fields, Eqs.~(\ref{erw}) and (\ref{e2rw}) are symmetric, while the third, \eref{egw}, is asymmetric. This point is important because for the fields $E_R(t)$ and $E_{2R}(t)$ the carrier frequency $\omega_0$ is equal to $\omega_1$, while for the field $E_G(t)$ we have $\omega_0\neq\omega_1$. Furthermore, the envelope of \eref{ert}, as well as the one of \eref{e2rt}, becomes the $\sinc$ function. Finally, for \eref{egt} the envelope results in a Gaussian function multiplied by the function $U(t)$, that is composed as a sum of two different imaginary error functions $\mathrm{erfi}(x)$.

The fields defined above represent three different situations. The field $E_R(t)$ is the most common expression for an ultrashort pulse, considering its spectral content is \textit{continuous}. The spectral function $\mathrm{rect}(\frac{\omega-\omega_1}{\Delta\omega})$ has the advantage of possessing a limited bandwidth, given by $\Delta\omega/2<\omega_1$, which prevents the pulse having spectral content near zero-frequency. This is particularly relevant in the few-cycle regime, where other envelopes typically used, e.g. Gaussian or $\sech$, are unable to fulfill this requirement. Near-zero frequencies are correlated with the ``zero-area pulse problem'' and are incompatible with the paraxial approximation used to focus the laser beams. In the $E_{2R}(t)$ field we observe the so-called frequency beat, considering we are summing up two carriers with frequencies separated by $\delta\omega$. These pulses have been already implemented in the laboratory, and possess interesting properties~\cite{he2019coherently,huang2011high,wirth2011synthesized}. The last one, $E_G(t)$, was chosen as a typical example of a field with an asymmetric spectrum and it is a relevant example where to test our hypothesis, since its carrier frequency $\omega_0$ is bandwidth dependent.

In Figs.~\ref{thpulses}(a)-\ref{thpulses}(c) we plot the spectral power $S(\omega)$ of the fields defined in Eqs.~(\ref{erw})-(\ref{egw}), for $E_0=1$. All the cases are centered at the same frequency, $\omega_1=2\pi$. The spectral bandwidth of both $E_R(\omega)$ and $E_G(\omega)$ is $\Delta\omega=\pi$, meanwhile for $E_{2R}(\omega)$ each `sub-spectrum' has a bandwidth $\Delta\omega=\pi/2$. For the latter $\delta\omega=\pi/2$. The temporal counterparts, $E_R(t)$, $E_{2R}(t)$ and $E_G(t)$, are depicted in Figs.~\ref{thpulses}(d)-\ref{thpulses}(f), where the thick blue (thin red) line defines a sine(cosine)-like pulse. 

%\begin{figure*}
%{\setlength{\tabcolsep}{-.5mm}
%\newlength{\spacfig}
%\setlength{\spacfig}{-7mm}
%\begin{tabular}{rl}
%\subfigure{\label{exppulses-a} %
%\includegraphics[width=0.5\textwidth]{Fig3a.png} } & 
%\subfigure{\label{exppulses-e} %
%\includegraphics[width=0.5\textwidth]{Fig3e.png} }\\[\spacfig]
%
%\subfigure{\label{exppulses-b} %
%\includegraphics[width=0.5\textwidth]{Fig3b.png} } & 
%\subfigure{\label{exppulses-f} %
%\includegraphics[width=0.5\textwidth]{Fig3f.png} }\\[\spacfig]
%
%\subfigure{\label{exppulses-c} %
%\includegraphics[width=0.5\textwidth]{Fig3c.png} } & 
%\subfigure{\label{exppulses-g} %
%\includegraphics[width=0.5\textwidth]{Fig3g.png} }\\[\spacfig]
%
%\subfigure{\label{exppulses-d} %
%\includegraphics[width=0.5\textwidth]{Fig3d.png} } & 
%\subfigure{\label{exppulses-h} %
%\includegraphics[width=0.5\textwidth]{Fig3h.png} }
%\end{tabular}
%}
%\caption{Experimental pulses.}
%\label{exppulses}
%\end{figure*}

Through the definition of the principal frequency $\omega_P$, \eref{e3}, we determine the principal period $T_P$, $T_P=2\pi/\omega_P$. In this way, we can analyze how the period $T$ of the pulses $E_R(t)$, $E_{2R}(t)$ and $E_G(t)$, represented as twice the distance between two adjacent maxima and minima in the temporal central region, is related to $T_P$ and the changes of the bandwidth $\Delta\omega$. In Fig.~\ref{thpulses}(d) we show how $T_P/2$ defines more precisely the temporal distance between two adjacent maxima and minima (exemplified for the $\cos$-like pulse), meanwhile $T_0/2$ is better suitable for the position of the field zeros (exemplified for the $\sin$-like pulse). 

\begin{figure}
{\setlength{\tabcolsep}{-.2mm}
\newlength{\spacfigv}
\setlength{\spacfigv}{-5mm}
\subfigure{\label{deltaw-a} %
\includegraphics[width=0.475\textwidth]{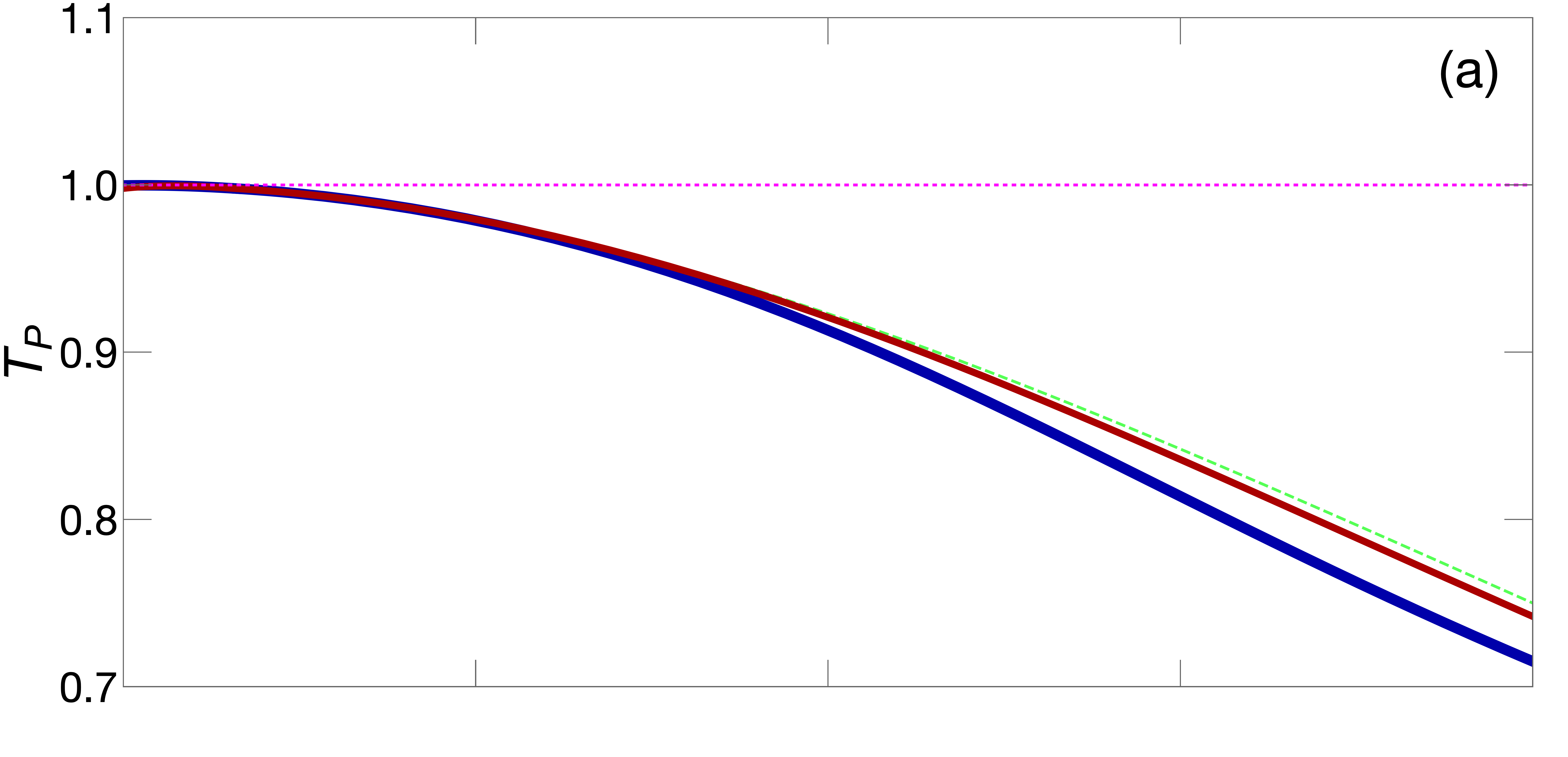} }\\[\spacfigv]
\subfigure{\label{deltaw-b} %
\includegraphics[width=0.475\textwidth]{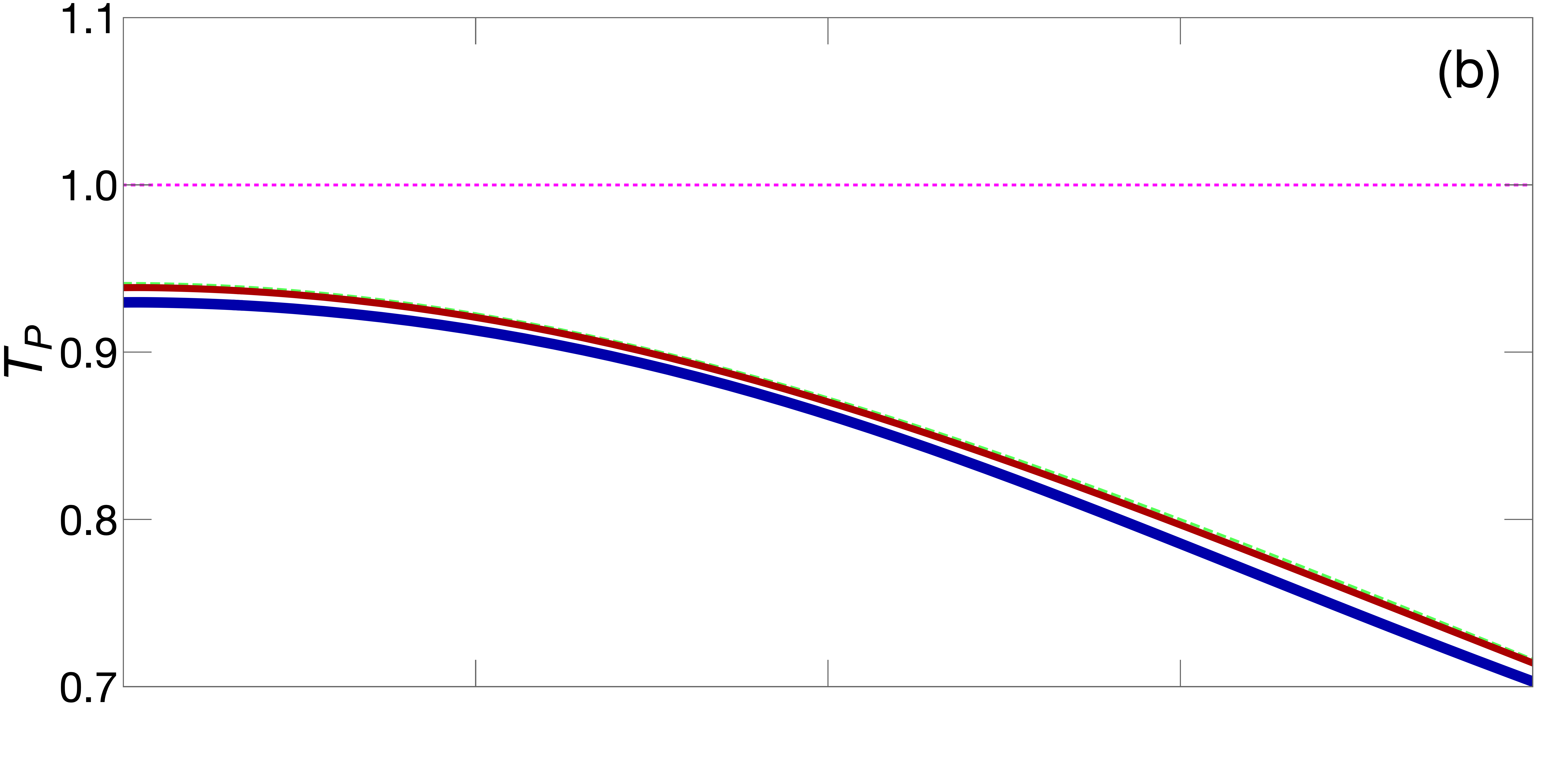} }\\[\spacfigv]
\subfigure{\label{deltaw-c} %
\includegraphics[width=0.475\textwidth]{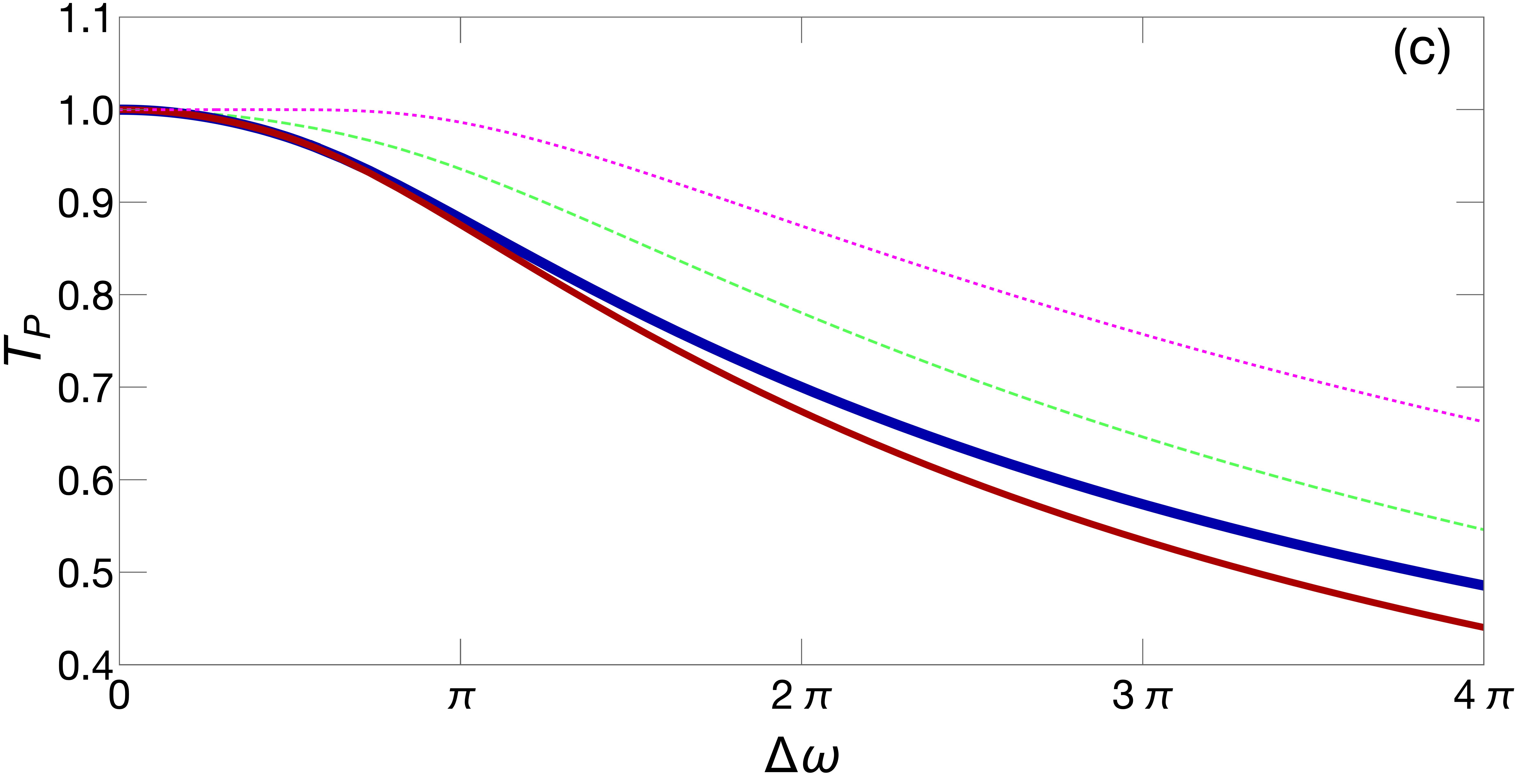} }
    \caption{Principal period $T_P$ as a function of the pulse bandwidth $\Delta\omega$ for (a) $E_R(t)$, (b) $E_{2R}(t)$ and (c) $E_G(t)$. In all the cases, dotted violet line: $T_0=2\pi/\omega_0$ (note that for the case of $E_G(t)$ pulses, $\omega_0$ depends on $\Delta\omega$), dashed green line: $T_P$ computed from $T_P=2\pi/\omega_P$ and thick blue (thin red) solid line: $T_{\mathrm{COS}}$ ($T_{\mathrm{SIN}}$) extracted for the time-dependent cosine(sine)-like pulses (see the text for details).}
    \label{fig2}
    }
\end{figure}

Let us now compute how $\omega_P$ changes as a function of the bandwidth $\Delta\omega$. For $E_R(\omega)$ is possible to find a simple analytical expression, meanwhile for both $E_{2R}(\omega)$ and $E_G(\omega)$ we deal with its numerical calculation. For the case of $E_R(\omega)$ we have:
%Because a similar result was showed in ~\cite{li2010carrier}, where a few-cycle rf pulse interacts with a two level system, we will show how is the analytic expression of $\omega_P$ for the spectrum $E_R(t)$, through of its definition \eref{e3}.
\begin{equation}
\label{e6}
\begin{split}
\omega_P(\Delta\omega,\omega_0) &=\frac{\int_{-\infty}^{\infty}\omega^2 S(\omega) \rmd\omega}{\int_{-\infty}^{\infty}\omega S(\omega) \rmd\omega}\\
&=\omega_0+\frac{\Delta\omega^2}{12\omega_0}.
\end{split}
\end{equation}
This last relationship between $\omega_P$, $\Delta\omega$ and $\omega_0$ (note that for this case $\omega_1=\omega_0$) has been shown in~\cite{li2010carrier}, for the case where a few-cycle RF pulse interacts with a two-level system.

In Fig.~(\ref{fig2}) we show how the principal period $T_P$ changes as function of the bandwidth $\Delta\omega$. The procedure to find the values of $T_P$ from the different cases is as follows. Starting at $t=0$ we search the position of a nearest minimum. If we define as $t_M$ the time where this minimum is located, for a cosine-like pulse we can compute the period $T_{\mathrm{COS}}$ as $T_{\mathrm{COS}}=2t_M$. Likewise, for a sine-like pulse becomes $T_{{\mathrm{SIN}}}=4t_M$. Figure \ref{fig2}(a) depicts the results for $E_R(t)$. The dashed green line represents $T_P$ computed as $T_P=2\pi/\omega_P$, meanwhile the thick blue (thin red) solid line corresponds to the $T_{{\mathrm{COS}}}$ ($T_{{\mathrm{SIN}}}$) extracted for the cosine(sine)-like pulses using the procedure explained above. We include the value $T_0=2\pi/\omega_0$ (dotted violet line), which is constant and equal to the unity for this case because the spectrum $E_R(\omega)$ is symmetrical. We can observe that $T_P$ appears to be a much more reliable quantity to predict the temporal distance between maxima and minima, for both cos- and sine-like pulses, and in a broad range of bandwidths. In \reffig{fig2}(b) we plot the results for $E_{2R}(t)$. Here, as well, we find an excellent agreement between the maxima and minima positions predicted by $T_P$ and those computed directly from the time-dependent field. Interestingly, for $\Delta\omega=0$, where the field results a sum of two continuous waves with frequencies $\omega_0-\delta\omega$ and $\omega_0+\delta\omega$, $T_P$ allows us to accurately find the positions of the maxima and minima. Finally, in \reffig{fig2}(c), we illustrate the case of $E_G(\omega)$. In this example, $T_0$ changes mainly due to the wide bandwidth of $E_G(\omega)$. Nevertheless, the positions of the maxima and minima computed using $T_P$ are much closer to those extracted from the fields than the ones calculated starting from $T_0$. We should note, however, that the agreement is not so good as in the previous cases. For asymmetrical pulses, as is this case, it is possible to show that an improvement in the maxima and minima positions prediction can be achieved by changing $S(\omega)$ by $S(\omega)^{1/2}$ in the definition of $\omega_P$, \eref{e3}, although this choice is relatively difficult to justify from first principles.

\begin{figure*}
\includegraphics[width=1\textwidth]{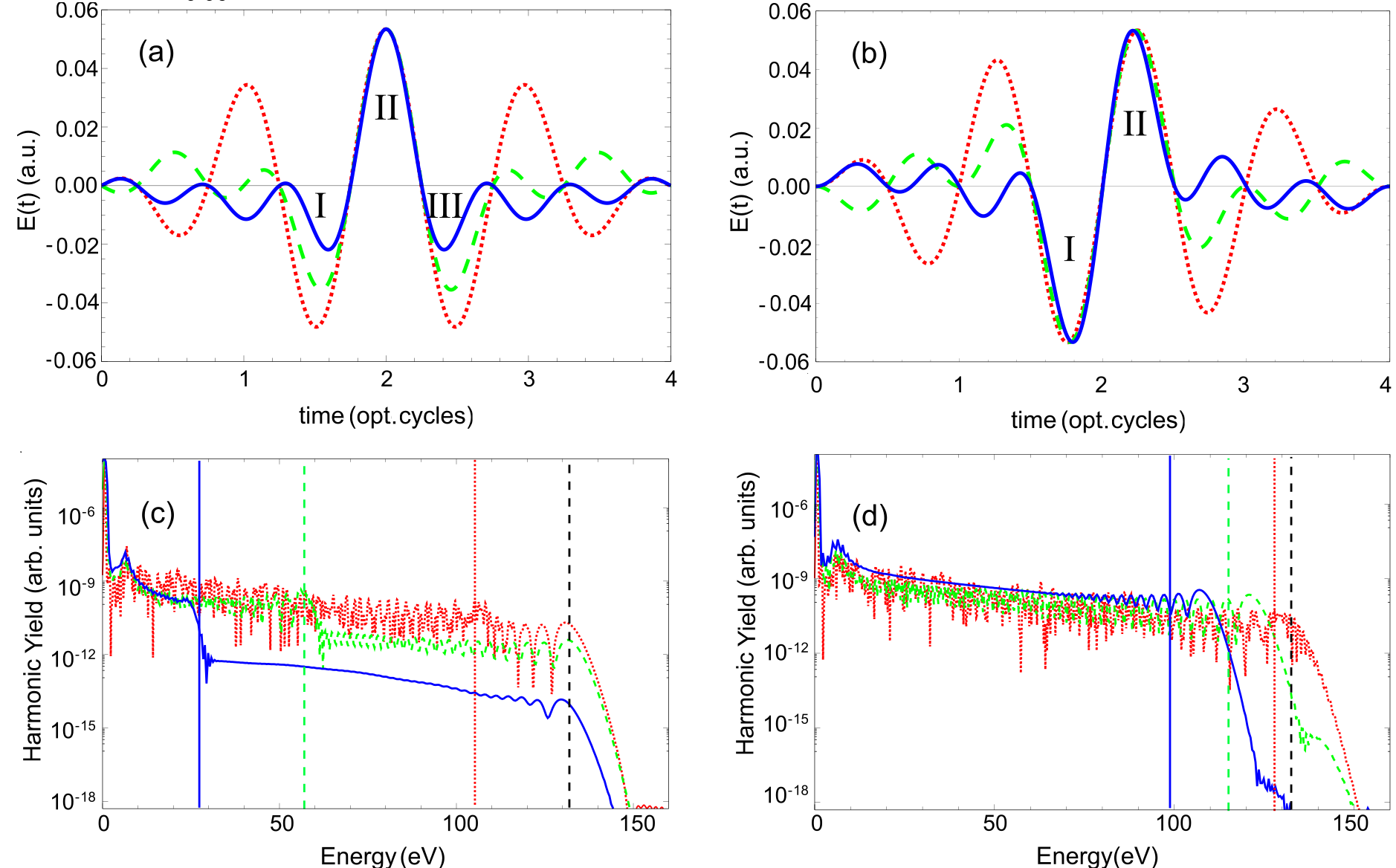}
\caption{Cosine-like (a) and sine-like (b) $E_R(t)$ pulses for different bandwidths $\Delta\omega$. The dotted red, dashed green and solid blue lines correspond to pulses with a bandwidth of $\Delta\omega=\pi$, $\Delta\omega = 2\pi$ and $ \Delta\omega = 3\pi$ in units of $\omega_0$, which we take corresponding to a laser wavelength of $\lambda_0$=2000 nm, respectively. The labels I-III denote the ionization and recombination regions (see the text for more details). HHG spectra for the Cosine-like (c) and Sine-like (d) pulses plotted in panels (a) and (b), respectively. The vertical lines correspond to the different HHG cutoffs (see the text for more details).\label{fig3}}
\end{figure*}

\section{Results and Discussion} 
\label{interaction}
 In this section we use some of the pulses previously described to drive an atomic system and characterize the high-order harmonic generation (HHG) spectra in terms of the principal frequency $\omega_P$. As examples, we employ both the $E_R(t)$ and $E_{2R}(t)$ fields to generate HHG in an hydrogen atom and simulate the dynamics through the numerical integration of the one-dimensional time-dependent Schr\"odinger equation (1D-TDSE). Recently, a theoretical investigation using sinc-shaped pulses for both HHG and the construction of a single attosecond pulse was presented~\cite{rajpoot2020}.

\subsection{$E_R(t)$ fields} 

To perform the numerical simulations, three values of the bandwidth $\Delta\omega$ were chosen for the $E_R (t)$ fields, namely $\Delta \omega = \pi$, $\Delta\omega=2\pi$ and $\Delta\omega = 3\pi$ in units of $\omega_0$, which we take corresponding to a laser wavelength of $\lambda_0=2000$ nm, both for the sine- and cosine-like fields. Taking into account the definition of the full-width at half-maximum (FWHM), we can find its value starting from the fields expression, Eq.~(\ref{ert}), as: FWHM$= 5.564/ \Delta\omega $. In this way, the corresponding FWHM result 1.77 opt.~cycles, 0.885 opt.~cycles. and 0.59 opt.~cycles, for $\Delta \omega = \pi$, $\Delta\omega=2\pi$ and $\Delta\omega = 3\pi$, respectively.

For all cases, the peak amplitude of the field was held fixed at $E_0 = 0.053$ a.u., which corresponds to a laser intensity of $ I =1\times 10^{14}$ W/cm$^2$. For cosine-like pulses, this value is reached in the central part of the pulse and is independent of the bandwidth. On the contrary, for sine-like pulses, the maximum amplitude of the field decreases, relative to the maximum of the envelope, as the pulse duration gets shorter; therefore, in order to keep the maximum value of the field at $ E_0 = 0.053$ a.u., we multiply the field amplitude by different scaling factors. In this way, taking into account that the HHG cutoff scales as $I\lambda^2$, any decrease in the maximum harmonic photon energy it is due to a change in the pulse wavelength or frequency and not to the peak amplitude of the field, product of the temporary shortening. In all the cases, we use a laser wavelength $\lambda_0 = $ 2000 nm and a hydrogen atom as a target ($ I_P = 0.5$ a.u.).

The different cos- and sine-like pulses are plotted in Figs.~\ref{fig3}(a) and \ref{fig3}(b), respectively. The dotted red, dashed green and solid blue lines correspond to pulses with a bandwidth of $\Delta\omega=\pi$, $\Delta\omega = 2\pi$ and $ \Delta\omega = 3\pi$, respectively. Taking into account Eq.~(\ref{e6}), the principal frequency of these pulses takes the following values: $\omega_P(\pi) = 2\pi(1 + 1/48)\approx 1.02 \omega_0$, $\omega_P (2\pi) = 2 \pi (1 + 1/12)\approx 1.08\omega_0 $ and $\omega_P (3\pi) = 2 \pi (1 + 3/16) \approx 1.19 \omega_0$. For these $\omega_P$ the associated wavelengths, thus, result $ \lambda_P(\pi)=\lambda_0 /1.02$, $\lambda_P(2\pi)=\lambda_0/1.08$ and $\lambda_P(3 \pi) =\lambda_0/1.19 $, respectively.

\begin{figure*}
\includegraphics[width=1\textwidth]{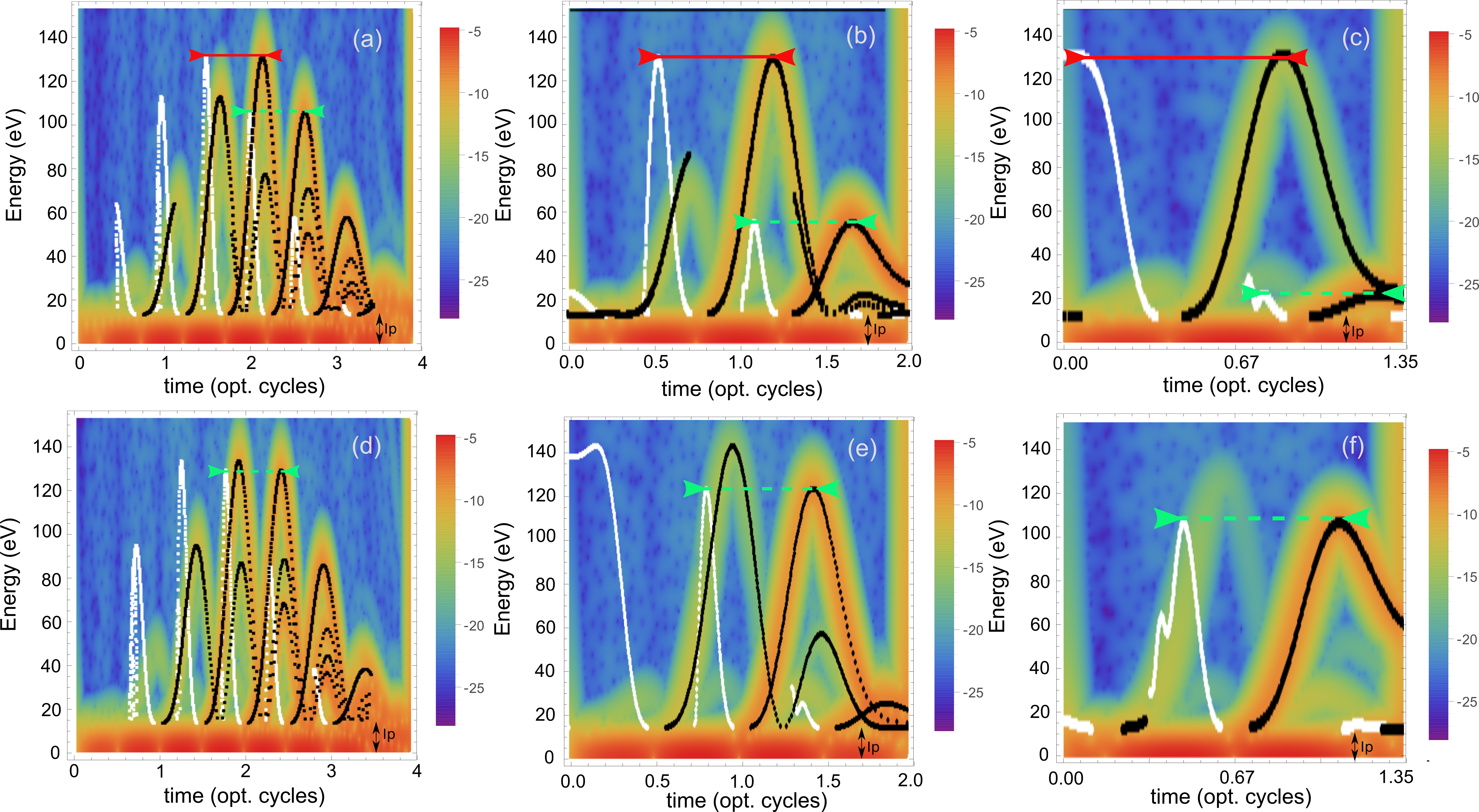}
\caption{Time-frequency  analysis  extracted from the 1D-TDSE HHG spectra. Superimposed we plot the classically computed rescattering energies of electrons as a function of the ionization time, in white dots, and recombination time, in black dots, for the laser pulses of Fig.~\ref{fig3}(a) (panels (a)-(c)) and Fig.~\ref{fig3}(b) (panels (d)-(f)).  In panels (a)-(c) the solid red (dashed green) arrow corresponds to the electron trajectory that contributes to the more (less) energetic HHG cutoff. Meanwhile, in panels (d)-(f) the green arrow corresponds to the electron trajectory that contributes to the single HHG cutoff (see the text for more details). The ionization potential $I_p$ energy is indicated by a black arrow.
\label{fig4}}
\end{figure*}

Figures~\ref{fig3}(c) and~\ref{fig3}(d) show the respective harmonic spectra for the pulses of Figs.~\ref{fig3}(a) and \ref{fig3}(b), obtained solving the 1D-TDSE (for details about the numerical implementation see e.g.~\cite{ciappina2012}).
The HHG for cosine-like pulses (Fig.~\ref{fig3}(c)) show two distinguishable plateaus in all the cases, with their corresponding cutoffs. The most energetic one, which corresponds to an energy given by $ I_0\lambda_0^2$, is marked with a dashed black line. Let us note that for all the cases this cutoff has the same photon energy (around 130 eV). On the contrary, the photon energy values for the less energetic cutoffs depend on the different pulses and decrease as the pulses become temporarily narrower.

 To explain this behavior, we perform a time-frequency analysis of the quantum mechanical computed HHG spectra. This tool has proven to be
very powerful to estimate the radiation emission times in
atoms and molecules and to discriminate the different classical electron trajectories. Furthermore, in this way we are able to visualize which temporal regions of the pulses correspond to the HHG cutoff. In short, starting with the quantum dipole acceleration $a(t)$, we apply the so-called Gabor transform defined as
\begin{eqnarray}
a_G(\Omega,t)&=&\int dt' a(t')\frac{\exp[-(t-t')^2/2\sigma^2]}{\sigma \sqrt{2\pi}}\exp(i\Omega t'),
\end{eqnarray}
where the integration is usually taken over the pulse duration and $\sigma$ is chosen in a way to achieve an adequate
balance between the time and frequency resolutions (typically $\sigma=1/3\omega$, being $\omega$ the central frequency of the laser pulse). We then plot $|a_G(\Omega,t)|^2$ as a function of the time and electron energy $E_k$ (note that in atomic units, where $\hbar=1$, $E_k=\Omega$). The results are shown in Figs.~\ref{fig4}(a)-\ref{fig4}(c), where we use a log scale for $|a_G(\Omega,t)|^2$. Additionally, we superimpose the electron kinetic energies at the recombination time as a function
of the ionization (white dots) and recombination times (black dots) calculated classically, using the Newton's equations of motion (for more details see e.g.~\cite{ciappinacpc}).

From this analysis we observe that the most energetic cutoffs (represented with a red solid line with arrows) originate in the temporal region I-II of the pulse (see Fig.~\ref{fig3}(a)). This means that the electron is ionized in region I and recombines in region II. On the contrary, the less energetic ones originate in the region II-III of the pulse (see the green dashed line with arrows in Figs.~\ref{fig4}(a)-\ref{fig4}(c)), i.e.~the electron is ionized in region II and recombines in region III. Let us notice that the peak amplitude of the field that ionizes the atom in region I is smaller than the one that does so in region II. This makes the probability of ionization-recombination in region I-II lower than the one in region II-III,~\cite{neyra2016high}. Furthermore, the classical analysis allows us to see what is the excursion time of the electron in the continuum for these two different temporal regions, I-II and II-III.

In the case of sine-like fields, the features of the harmonic spectra are as follows. An extended plateau, with a clear and single cutoff, is observed in the different spectra (Fig.~\ref{fig3}(d)), which decreases as the pulse temporarily becomes shorter. The time-frequency analysis of these spectra is shown in Figs.~\ref{fig4}(d)-\ref{fig4}(f), respectively. For all the cases, the most probable ionization-recombination event, indicated by a green dashed line with arrows, occurs at the central part of the pulse, i.e.~the electron is ionized in region I and recombines in region II (see Fig.~\ref{fig3}(b)). Furthermore, these pulses are symmetric in the region of ionization and recombination (I and II), analogously to a continuous field.

The introduction of the principal frequency $\omega_P $, allow us a better interpretation of the results described above. In Fig.~\ref{thpulses}, we show that the period of time between a maximum and a minimum of the field in the central region of the pulse is accurately described by the period $T_P$, associated with the principal frequency $\omega_P$ ($T_P=2\pi/\omega_P$). We have also shown that $T_P$ decreases as the spectral content of the pulse increases (the FWHM value of the pulses gets smaller). This result is clearly visible in the fields represented by Figs.~\ref{fig3}(a) and \ref{fig3}(b), where we see that as the FWHM decreases, the period $T_P$ does so as well.

\begin{figure*}
\includegraphics[width=.85\textwidth]{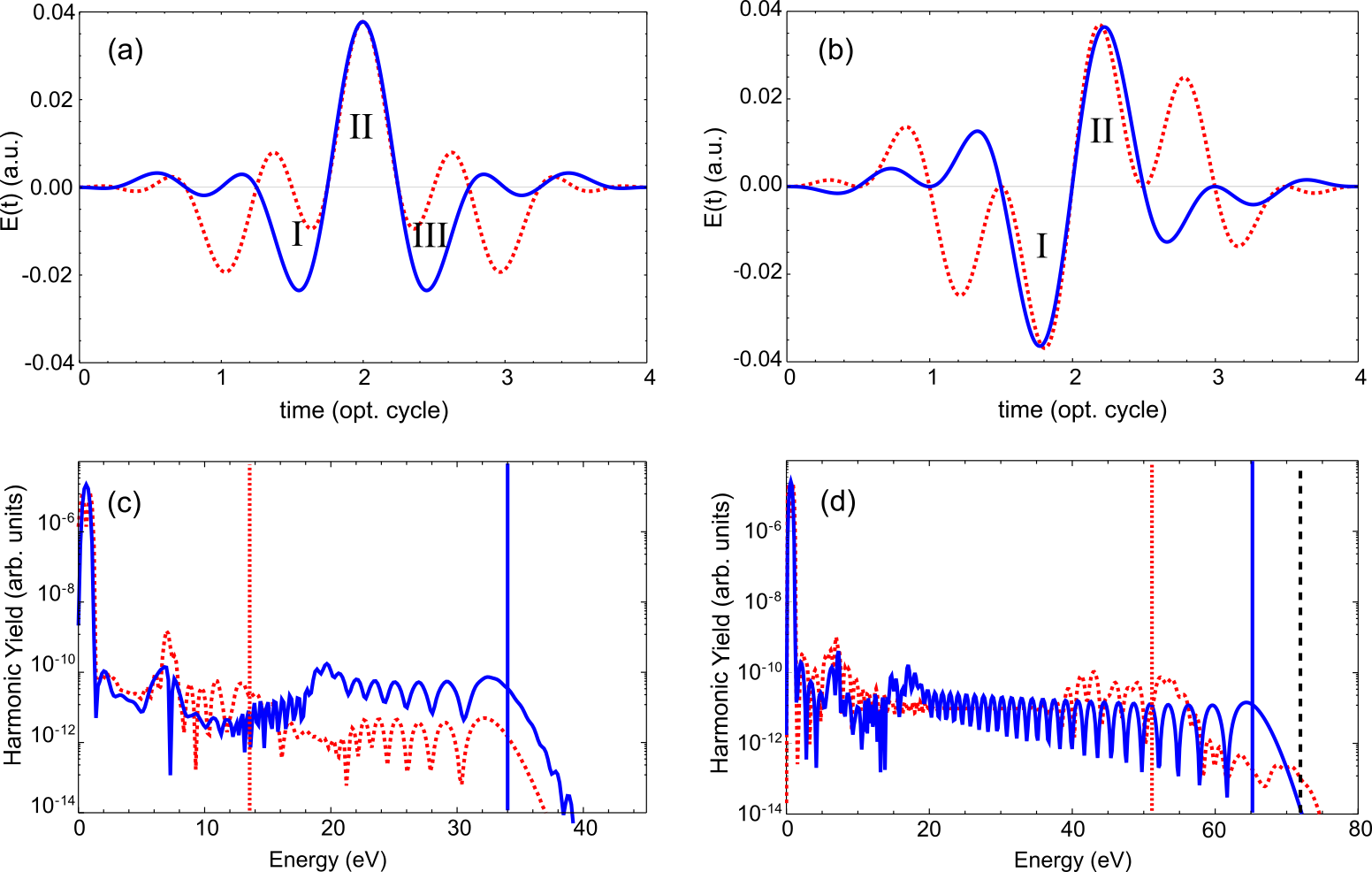}
\caption{Cosine-like (a) and sine-like (b) $E_{2R}(t)$ pulses for different values of $\Delta\omega$ and $\delta\omega$. The dotted red (solid blue) line corresponds to pulses with $\Delta\omega=0.1$ and $\delta\omega=\pi$ ($\Delta\omega = \pi/2$ and $\delta \omega = \pi/2 $). The labels I-III denote the ionization and recombination regions (see the text for more details). HHG spectra for the cosine-like (c) and sine-like (d) pulses plotted in panels (a) and (b), respectively. The vertical lines correspond to the different HHG cutoffs (see the text for more details).\label{fig5}}
\end{figure*}

For the case of the spectra generated by the sine-like fields (Fig.~\ref{fig3}(d)), the reduction in $T_P$ is observed as a decrease in the cutoff, as the pulse becomes temporarily shorter. Furthermore, the electron time of flight in the continuum $\tau$, for the most energetic trajectories, decreases accordingly. This can be seen from the classic analysis shown in Figs.~\ref{fig4}(d)-\ref{fig4}(f), where the times of flight, denoted by green dashed arrows, are: $\tau(\pi)\approx 0.64$ opt.~cycles, $\tau(2\pi)\approx 0.627$ opt. ~cycles and $\tau(3\pi)\approx 0.6 $ opt. cycles.~This analysis is relevant considering the efficiency of the harmonic generation depends on the time of flight of the electron in the continuum and scales as $\propto\lambda$ ~\cite{haessler2014optimization}.
To calculate the classical cutoff of the different spectra, as a function of the frequency $\omega_P$, we use the formula $I_P + 3.17U_P$,where $U_P$ is the ponderomotive energy, $U_P=\frac{E_0^2}{4\omega_P^2}$, and $I_P$ the ionization potential. Thus, for $\lambda_0=2000$ nm, we get: $\lambda_P(\pi)=1960$ nm, $\lambda_P(2\pi)=1850$ nm and $\lambda_P(3\pi)=1680$ nm, which correspond to HHG cutoff at $\approx 127$ eV, 115 eV and 97 eV, respectively. These values are indicated by the dotted red, dashed green and solid blue lines in Fig.~\ref{fig3}(d), respectively.

In the case of cosine-like pulses the situation is different. As mentioned before, there are two temporal regions in the pulses that contribute to the harmonic spectrum cutoff. The region I-II, for all cases, generates a cutoff  given by $I_0\lambda_0^2$ and has a value of approximately 132 eV. In contrast, the region II-III of the pulse generates a cutoff that depends on $\Delta\omega$. If we suppose that the kinetic energy the electron acquires in the continuum is given by the region of the pulse that recombines it, in this case region III~\cite{haworth2007half,luke2009}, the intensity $I_0$ must be adjusted as a function of the field amplitude in such temporal region, in order to keep the HHG cutoff $\propto I\lambda_0^2$. Thus, we obtain the following intensities: $ I (\pi) =0.9^2\cdot I_0$, $ I(2\pi) =0.656^2\cdot I_0$ and $ I(3\pi) = 0.41^2\cdot I_0$ (being the relative amplitude of the fields in region III equal to 0.9, 0.656, 0.41, respectively). If we also assume that the frequency dominating this region is $\omega_P$, we obtain the following cutoffs: 105.6 eV, 57 eV and 27.6 eV. The position of these cutoffs is denoted in Fig.~\ref{fig3}(c) with a dotted red, dashed green and solid blue line, respectively. As can be observed, these values for the less energetic cutoff are in very good agreement with the 1D-TDSE predictions. Therefore, for the case of cosine-like pulses, the laser-matter interaction seems to be dominated simultaneously by a first region (I-II), governed by the carrier frequency $\omega_0$, and another region (II-III) governed by the principal frequency $\omega_P$ ~\cite{ishii2014carrier}.

\subsection{$E_{2R}(t)$ fields} 

For the simulations performed with the pulses $E_ {2R}(t)$, two representative examples were chosen, based on the parameters $\Delta\omega$ and $\delta\omega$ (see Eqs.~\ref{e2rt}(b) and~\ref{e2rw}(b)). The respective laser electric fields are shown in Figs.~\ref{fig5}(a) and~\ref{fig5}(b) (for the cosine-like and sine-like pulses, respectively), where the dotted red (solid blue) line denotes the pulse with $\Delta\omega=0.1$ and $\delta\omega=\pi$ ($\Delta\omega=\pi/2$ and $\delta\omega=\pi/2$).

As in the case of the $E_{R}(t)$ fields, the central wavelength was set to $\lambda_0=$ 2000 nm and the target is a hydrogen atom with $I_P= 0.5$ a.u. The peak electric field amplitude is now $E_0=0.037$ a.u., which corresponds to a laser peak intensity of $ I_0= 5\times 10^{13}$ W/cm$^ 2 $. Analogously to the case of the $E_R (t)$ fields, for the sine-like pulses, the peak field amplitudes are multiplied by a scale factor, in such a way that their values are $ E_0 = 0.037$ a.u.~in all the cases.

\begin{figure*}
\includegraphics[width=.85\textwidth]{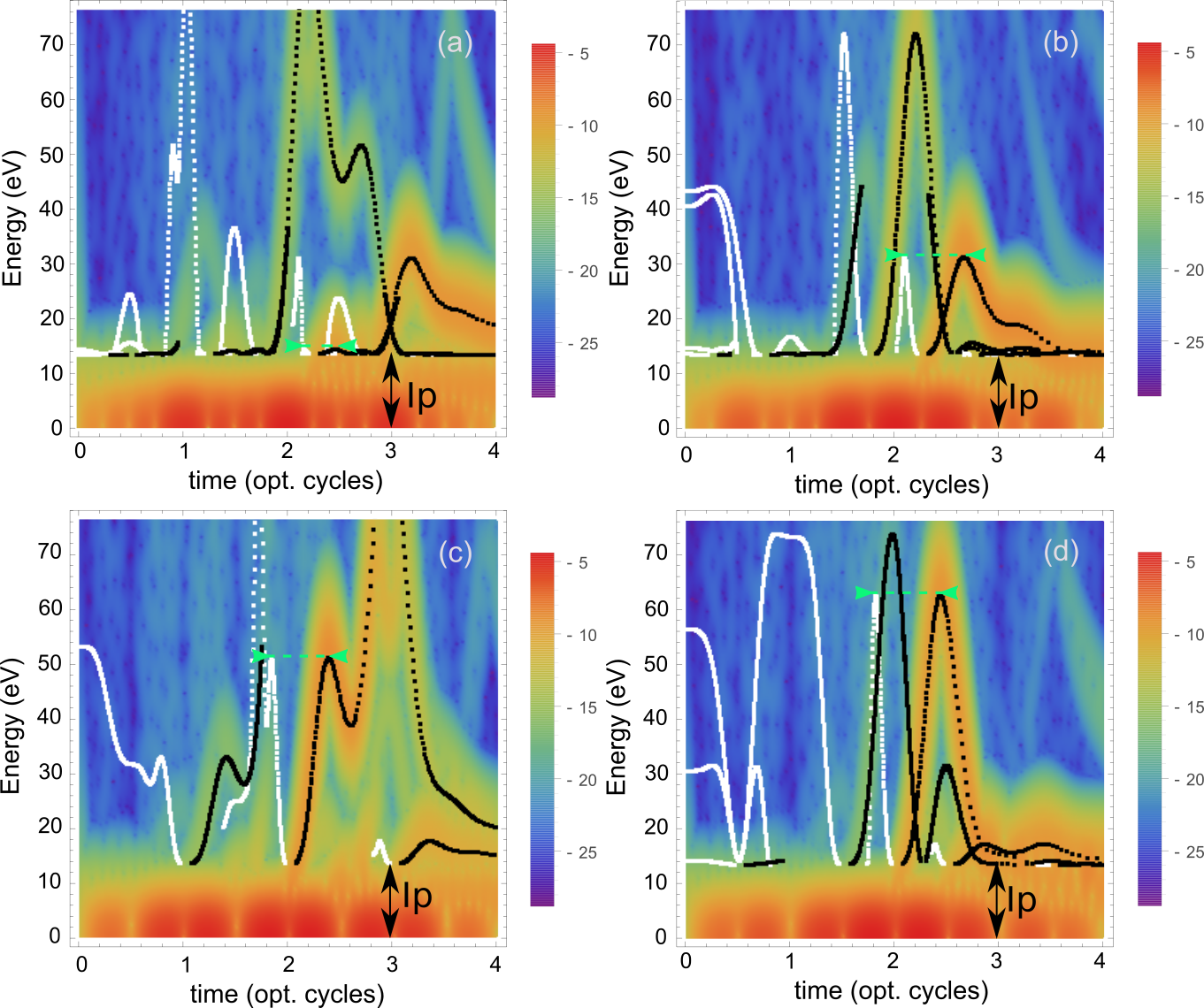}
\caption{Time-frequency analysis extracted from the 1D-TDSE HHG spectra. Superimposed we plot the classically computed rescattering energies of electrons as a function of the ionization time, in white dots, and recombination time, in black dots, for the laser pulses of Fig.~\ref{fig5}(a) (panels (a) and (b)) and Fig.~\ref{fig5}(b) (panels (c) and (d)).  In all the panels the green arrow corresponds to the electron trajectory that contributes to the single HHG cutoff (see the text for more details). The ionization potential $I_p$ energy is indicated by a black arrow. \label{fig6}}
\end{figure*}

Figures~\ref{fig5}(c) and~\ref{fig5}(d) show the harmonic spectra generated by the pulses given by Figs.~\ref{fig5}(a) and~\ref{fig5}(b), respectively.
In the case of sine-like pulses, the structure of the harmonic spectrum is analogous to that shown for the $ E_R (t) $ pulses. That is, there is a continuous spectrum of harmonics, which reaches a well-defined maximum value. In the case of cosine-like pulses the situation is different. The pulse represented by a solid blue line ($\Delta\omega = \pi/2 $, $\delta\omega=\pi/2$), generates a continuous spectrum similar to a sine-like pulse. On the contrary, for the pulse represented by the dotted red line ($ \Delta\omega = 0.1 $ and $\delta\omega = \pi $), it can be seen that there are two plateaus with their respective cutoffs, at energies around 15 eV and 35 eV. To analyze these structures in the spectra and relate them to the principal frequency defined in Eq.~(\ref{e3}), we carried out a time-frequency analysis (see Fig.~\ref{fig6})), on which, in addition, we superimpose the classically computed rescattering energies of electrons as a function of the ionization time, in white dots, and recombination time, in black dots. Figs.~\ref{fig6}(a) and~\ref{fig6}(b), correspond to the cosine-like pulses, dotted red and solid blue lines, respectively (see Fig.~\ref{fig5}(a)) and Fig.~\ref{fig6}(a). Here we see how the electron ionization probability is greater in region II of the pulse and practically zero in region I. This is due to the fact that the amplitude of the field in region I is not intense enough as to ionize the electron by tunnelling. In this way, the dominant event occurs when the electron is ionized in region II and recombines in region III. This event is shown with a green dashed line with arrows and corresponds to a maximum energy of approximately 15 eV. In Fig.~\ref{fig6}(b), a similar situation is shown: the ionization probability in region I of the pulse is negligible, in relation to the one in region II. In this way there is only one ionization-recombination event, which takes place in the temporal region II-III of the pulse. This event is shown with a green dashed line with arrows and has a maximum energy of approximately 30 eV.

For sine-like pulses, the analysis is shown in Figs.~\ref{fig6}(c) and~\ref{fig6}(d) (dotted red and solid blue lines, respectively). For these cases, we observe there is a predominant ionization-recombination event, which originates in the region I-II of the pulse, analogously to the case of the pulses $E_R (t)$.

To explain the spectra obtained as a function of the principal frequency $\omega_P$, we proceed to compute its value for the two pulses presented. Using the definition, Eq.~(\ref{e3}), we obtain $\omega_P =1.25 \omega_0$ for the case $\Delta\omega = 0.1 $ and $\delta\omega=\pi$ and $ \omega_P=1.06\omega_0$ for the case $\Delta\omega=\pi/2$ and $\delta\omega = \pi/2$. Taking into account that for the case of cosine-like pulses the peak field amplitude relative to the maximum amplitude in region III is: 0.26 and 0.62 (dotted red and solid blue solid lines, respectively), we proceed to calculate the different cutoffs in the same way as for the case of the pulses $E_R(t)$. These are indicated with a dotted red and solid blue line in Figs.~\ref{fig5}(a) and~\ref{fig5}(b), where the cutoff for an intensity of $ I_0 = 5\times 10^{13}$ W/cm$^2$ and a wavelength of $\lambda_0 = 2000$ nm is indicated by a black dashed line. For sine-like pulses with $ \omega_P (0.1, \pi) = 1.25\omega_0 $, the cutoff results  51.4 eV and with $ \omega_P(\pi/2, \pi/2) = 1.06 \omega_0 $, 66.2 eV. For cosine-like pulses, with $\omega_P (0.1, \pi) = 1.25 \omega_0$ we obtain a cutoff at 16 eV and with $ \omega_P(\pi/2, \pi/2) = 1.06 \omega_0 $ at 33.8 eV. These last values are in excellent agreement with the quantum mechanical simulations.

\section{Conclusions and Outlook}
\label{conclusions}

In conclusion, a new definition of the principal frequency of an ultra-short laser pulse is introduced. This new frequency, called principal frequency, describes in a better way the interaction between an ultrashort pulse with matter, particularly when the spectral content of the pulse has more than one octave, which in the temporal domain corresponds to the single- or sub-cycle regime.

In addition, through the principal frequency $\omega_P$ the CEP effects in the HHG can be interpreted in an alternative way. For sine-like pulses, a single HHG cutoff is generated by a single ionization-recombination event (one region of recombination associated to one region of ionization) between two symmetric regions at the center of the pulse, in which the principal frequency $\omega_P$ dominates the interaction.  On the other hand, for cosine-like pulses, there are two ionization-recombination events that contribute to the development of several HHG cutoffs. The first ionization-recombination event is governed by the carrier frequency $\omega_0$, in the region in which the amplitude of the field that ionizes the atom is smaller that the maximum peak field amplitude,  responsible to the recombination. The second ionization-recombination event is dominated by the principal frequency $\omega_P$. Here, the ionization of the atom takes place in the region of the pulse with higher peak amplitude and the recombination in the region smaller peak amplitude. 

These two events are clearly visible on the HHG spectra as two plateaus, with different cutoff energies and efficiency. This is so because the cutoff scales as $\propto \lambda^2$ and the trajectory excursion time duration as $\propto\lambda$.  Because in the central part the $E_G(t)$ pulses have a temporal shape  similar to the $E_R(t)$ ones, results obtained with this kind of pulses do not contribute to the final discussion. We have noted, however, that for pulses with an asymmetric frequency spectrum or a more complex frequency contents, a definition of the principal frequency $\omega_P$ using a density $\tilde{\rho}(\omega)=\omega S(\omega)^{1/2}$ gives better predictions for HHG cutoff. On the other hand, the latter definition would be hard to justify from first principles.

The introduction of the principal frequency $\omega_P$ suggests that the distribution of the photons "inside" of an ultra-short pulse is not linear. We show that the principal frequency is shifted to the higher frequencies when the spectral width of the pulse increases. On the other hand, a similar result was showed in the temporal domain through the introduction of an "intrinsic chirp" in THz pulses~\cite{lin2006subcycle, lin2010intrinsic, ruchert2013spatiotemporal}, particularly when those pulses are focused.

As the non-linear response of matter depends on some power of the electric field amplitude ($E_0^ n$), it is to be expected that for few-cycle pulses, the interaction is dominated by the principal frequency $\omega_P$ (or period $T_P$), which is the one that defines the position of the maxima and minima of the field~\cite{rybka2016sub}.

\section*{Acknowledgements}

ICFO group acknowledges support from ERC AdG NOQIA, Spanish Ministry of Economy and Competitiveness (``Severo Ochoa'' Program for Centres of Excellence in R\&D (CEX2019-000910-S), Plan National FISICATEAMO and FIDEUA PID2019-106901GB-I00/10.13039/501100011033, FPI), Fundaci\'o Privada Cellex, Fundaci\'o Mir-Puig, and from Generalitat de Catalunya (AGAUR Grant No.~2017 SGR 1341, CERCA program, QuantumCAT\_U16-011424, co-funded by ERDF Operational Program of Catalonia 2014-2020), MINECO-EU QUANTERA MAQS (funded by State Research Agency (AEI) PCI2019-111828-2 / 10.13039/501100011033), EU Horizon 2020 FET-OPEN OPTOLogic (Grant No 899794), and the National Science Centre, Poland-Symfonia Grant No. 2016/20/W/ST4/00314.

\listoforcidids
 %M. F. C. sincerely thanks T. T. Luu for providing us his experimental data in editable format. 

%\begin{figure}[b]
%\includegraphics[width=0.45\textwidth]{Fig2.0.png}% Here is how to import EPS art
%\caption{\label{fig2a} A figure caption. The figure captions are
%automatically numbered.}
%\end{figure}

%\begin{figure}[b]
%\includegraphics[width=0.4\textwidth]{Fig3.0.png}% Here is how to import EPS art
%\caption{\label{fig3} A figure caption. The figure captions are
%automatically numbered.}
%\end{figure}

%\begin{figure}[b]
%\includegraphics[width=0.4\textwidth]{Fig4.0.png}% Here is how to import EPS art
%\caption{\label{fig4} A figure caption. The figure captions are
%automatically numbered.}
%\end{figure}

% Automatic bibliography
%\bibliographystyle{arthur} 
%\bibliography{references}{}

\end{document}